\documentclass[a4paper,11pt,
]{article}
\usepackage[OT2,T1]{fontenc}
\usepackage[utf8]{inputenc}
\usepackage{textcomp}

\usepackage{indentfirst}
\usepackage{microtype}
\usepackage[tbtags]{amsmath}
\usepackage{mathrsfs}
\usepackage{amssymb}
\usepackage[
noblocks,
max3,
]{authblk}
\usepackage{hyperref}
\makeatletter
\renewcommand*{\@fnsymbol}[1]{\ensuremath{\ifcase#1\or \text{\textreferencemark} \or *\or \dagger\or \ddagger\or \mathsection\or \mathparagraph\or \|\or **\or \dagger\dagger \or \ddagger\ddagger \else\@ctrerr\fi}}
\makeatother
\newcommand{\ton}[1]{\left(#1\right)}

\newcommand{\eq}[1]{\( #1 \)}

\newcommand{\Rom}[1]{\mathrm{#1}}
\title{\textbf{Between quantum and classical gravity: \\
Is there a mesoscopic spacetime?}}

\author[1,2]{Eolo Di Casola\thanks{E-mail: \url{eolo.dicasola@sissa.it}}}
\author[1,2]{Stefano Liberati\thanks{E-mail: \url{stefano.liberati@sissa.it}}}
\author[3]{Sebastiano Sonego\thanks{E-mail: \url{sebastiano.sonego@uniud.it}}}

\affil[1]{SISSA, via Bonomea 265, 34136 Trieste, Italy}
\affil[2]{INFN, Sezione di Trieste, Via Valerio 2, 34127 Trieste, Italy}
\affil[3]{DCFA, Sezione di Fisica e Matematica, Universit{\`a} di Udine, Via delle Scienze 206, 33100 Udine, Italy}

\date{\small November 28th, 2014}
\begin{document}

\maketitle

\begin{abstract}
Between the microscopic domain ruled by quantum gravity, and the macroscopic scales described by general relativity, there might be an intermediate, ``mesoscopic'' regime, where spacetime can still be approximately treated as a differentiable pseudo-Riemannian manifold, with small corrections of quantum gravitational origin.  We argue that, unless one accepts to give up the relativity principle, either such a regime does not exist at all --- hence, the quantum-to-classical transition is sharp ---, or the only mesoscopic, tiny corrections conceivable are on the behaviour of physical fields, rather than on the geometric structures.
\end{abstract}

\bigskip


\bigskip\medskip

``Spacetime'' is a powerful word in physics. It refers simultaneously to some ``grand stage'', where all phenomena can unfold, but also to a dynamical entity itself, governed by Einstein's equations.  Mathematically, it is an umbrella term encompassing both ``pre-physical'' structures (dimension, topology, differentiable structure etc., usually absorbed into the notion of an appropriate smooth manifold), and objects with a physical interpretation (the metric, the curvature, a temporal orientation, etc.)~\cite{grclassics}.

Yet, the same word also evokes a theoretical landscape far from being settled.  Our present model of spacetime as a pseudo-Riemannian differentiable manifold with Lorentzian signature, can be considered satisfactory and efficient from cosmological scales down to particle physics scales (energies of at least \eq{10^{20}} eV, from ultra-high-energy cosmic rays). However, quantum fluctuations emerging at microscopic level are expected to shatter the classical structure of space and time at small scales  (frequently, one refers to the Planck length \eq{\ell_{\mathrm P} \sim 10^{- 33}}~cm). This might imply changes in the physical fields and/or in the ``pre-physical'' structures, with a strong dependence on the model adopted~\cite{rovelli-loop,oriti-gft,dowker-causets}. 

In principle, any such microscopic variation of the classical scheme might propagate upwards to a ``transition'' scale \eq{\ell} much larger than the Planck one.  It could thus entail \emph{mesoscopic} modifications of spacetime~\cite{Sorkin:2007qi}, already within our present experimental window, where much effort is focussed to provide tight constraints~\cite{Liberati:2013xla}.  The picture is then the following: below \eq{\ell_{\Rom{P}}}, spacetime (or whatever replaces it) can no longer be  modelled after a pseudo-Riemannian differentiable manifold, which provides instead an accurate description at scales larger than \eq{\ell}.  In the mesoscopic region between \eq{\ell_{\mathrm P}} and \eq{\ell}, such model may still be viable, provided that appropriate (small) corrections be applied to the laws of physics and/or to the geometrical objects. Just to give one example, in the context of Causal Sets theory, \eq{\ell} is a non-locality scale bridging the gap between the microscopic causal network at the Planck scale, and the macroscopic, smooth spacetime manifold~\cite{Sorkin:2007qi}.

It is fair to ask, then, what could be the common features of the corrections expected to emerge in this mesoscopic regime. To begin with, current observations imply that the scale \eq{\ell} must be much smaller than any curvature radius associated with macroscopic gravitational fields. At the same time, it is much larger than the Planck length, where quantum gravitational effects become relevant.  These are exactly the conditions under which we would expect special relativity to hold.  Therefore, any mesoscopic-regime deviation from standard physics induced by the sub-Planckian  behaviour of spacetime, either can show up as corrections to the physical laws in ordinary Minkowski spacetime, or, more radically, can be ascribed to changes of the Minkowskian structure itself.  In order to understand which possibilities are there for the latter deviations, we shall then tinker with the founding pillars of special relativity.

\bigskip

It is worth stressing at this stage that the notion of ``spacetime structure'' admits two possible interpretations. One can believe that it reveals some set of underlying features, which surface through the dynamics of observers and measuring devices, but exist independently of these systems. This conclusion, however, is not mandatory, and one can also take a less elaborate view, in which ``spacetime structure'' is just a convenient notion to express suitable properties of measurements.  Space and time thus become mere bookkeeping devices, useful to organise such measurements, rather than physical entities on their own.  In this sense, whether there is or not a regime in which it makes sense to speak of a spacetime, depends on the feasibility (in principle) of these operations.  This is the view that lies, often implicitly, at the roots of most derivations of the kinematical transformations, and is the one we shall adopt in the following.

A postulate lying at the very foundations of classical, non-gravitational physics is that, in any given region of spacetime, one can find at least one system of observers, define procedures for synchronising clocks, and choose units for length, such that the distance between any two observers does not depend on time, and that such distance satisfies the Euclidean axioms.  Moreover, clocks are chosen to measure time in such a way that the most elementary laws of physics take their simplest form~\cite{mtw}.

These assumptions are seldom made explicitly, but are fundamental in the development of both Newton's and Einstein's mechanics.  We shall refer to them collectively as postulate (A). Among many other things, they allow us to describe events using the reference frames germane to all treatments of kinematics, i.e. an ordered set of coordinates \eq{x^a} (\eq{a = 1,2,3,4}, with \eq{x^4 = t}), with the following operational meaning: the differences \eq{\Delta x^a = x^a_Q - x^a_P} for some ordered pair of events \eq{\ton{P,Q}} are values of distances and durations measured by the observers.

Frames differing from each other by spatial rotations and translations, and by time translations, are physically equivalent --- they correspond to the same system of observers, for which (A) is supposed to hold.  Experience, however, suggests that such a system of observers is not unique: if (A) holds for a system of observers, then it also holds for any other system whose observers have all the same constant velocity with respect to the first system.  We shall call this postulate (B).

We can then consider the maps \eq{x'^a = f^a \ton{x^1,x^2,x^3,x^4; \boldsymbol{v}}}, relating the coordinates of an event in two frames adapted to the two systems of observers (\eq{\boldsymbol{v}} is the velocity of the second frame with respect to the first one).  It has been known for more than a century, starting from the pioneering work by von Ignatowsky, that the form of the maps \eq{f^a} can be derived assuming the following four hypotheses~\cite{Liberati:2013xla, brown2005, Sonego:2008iu}:
\begin{enumerate}
\item[(i)] Spatial and temporal homogeneity (equivalence of all locations in space, and instants of time);
\item[(ii)] Spatial isotropy (equivalence of all directions in space);
\item[(iii)] Principle of relativity (absence of a preferred frame --- mathematically, this property bestows a group structure on the functions \eq{f^a});
\item[(iv)] Pre-causality (the temporal ordering of events occurring at the same spatial location in one frame cannot be reversed in another frame).
\end{enumerate}
The result is the Lorentz transformation, containing an undetermined constant with the meaning of an invariant speed.  The latter might in principle be infinite (yielding the Galilei transformation), but experiments show that it coincides with the speed of light in vacuum. 

Summarising, the Minkowskian structure is a consequence of postulates (A) and (B), and of assumptions (i)--(iv).  Therefore, a mesoscopic regime exists where spacetime structure deviates from the special relativistic one, only if at least one of these postulates/assumptions does not hold.  This point is worth stressing, for Lorentz invariance is often identified with the relativity principle.  Actually, a violation of \emph{any} of the assumptions above would result in a departure from Lorentz invariance.  Let us then explore the various possibilities.

\bigskip

The most robust assumption in the list above is probably pre-causality, item (iv).  Its role is to identify \eq{t} as a variable expressing clock measurements (conceptually different from the measurements of distance expressed by the other three coordinates).  Thus, it is actually part of the characterisation of reference frames, and as such it cannot be relaxed within the operational approach.  A possible violation of pre-causality in the mesoscopic regime (keeping, of course, all the other assumptions) would thus indicate the inapplicability of the entire operational construction based on observers and clocks.  This would be the case, for instance, if a transition from Lorentzian to Euclidean signature were taking place.\footnote{One should remind, however, that if quantum field theory can still be applied in the presence of such a signature change, this is phenomenologically constrained by quantum instabilities~\cite{dray-signchange,silke-signchange}.}

One obvious option is to relinquish the principle of relativity, hypothesis (iii), which results in a vast catalogue of proposals~\cite{Baccetti:2011aa}. However, many eminent models of quantum/emergent gravity enforce this principle at the very fundamental level~\cite{rovelli-loop,oriti-gft,dowker-causets, Rovelli:2010ed}, and quite tight experimental constraints have been cast~\cite{Liberati:2013xla}.  

A breakdown of isotropy at a kinematical level --- assumption (ii) --- is fully compatible with the relativity principle and theoretically viable, provided that one trades the pseudo-Riemannian metric for a pseudo-Finslerian one, the manifold structure of spacetime remaining intact~\cite{Sonego:2008iu, Gibbons:2007iu}. Yet, the onset of a privileged direction in space is not the kind of effect one would expect to emerge from an underlying quantum regime.  

We are thus left, among assumptions (i)--(iv), only with item (i), spatial and temporal homogeneity.  The technical role of this hypothesis is to make the functions  \eq{f^a} linear in the \eq{x^a}.  Indeed, this is a very powerful implication: once linearity is established, no room is left for any constant with dimensions other than a speed.  It seems, then, that (i) is the ingredient to give up if one wants to recover some kind of scale dependence, with some fundamental length like \eq{\ell_\mathrm{P}} appearing in the \eq{f^a}.  Relaxing homogeneity/linearity, however, amounts to abandoning the operational interpretation of the coordinates \eq{x'^a}, for the differences \eq{\Delta x'^a} no more account for any duration and/or length in a given reference frame.  If the relativity principle (iii) is still supposed to hold, this is true for the coordinates in any frame.  But the possibility to set up coordinates with a straightforward operational interpretation is a very basic assumption, guaranteed by postulate (A).  It then results that any attempt to weaken homogeneity demands a radical redefinition of ground-level concepts.

\bigskip

We have shown that, if one assumes that spacetime in the mesoscopic regime still obeys postulates (A)--(B), plus hypotheses (i)--(iv), then it has a full Minkowskian structure, and it makes no sense at all to call it ``mesoscopic''.  In this case, if anything unforeseen crops up at a microscopic level, it does so abruptly, as in a second-order phase transition.

On the other hand, one might try to relax the very postulate (A).  Then, the operational construction of spacetime structure would not be possible anymore --- at least, not in the usual way.  (This would be the case, for instance, in a scenario containing nonlocal defects~\cite{Hossdefects}, if the latter are assumed to alter the worldlines of observers.)  If, under these conditions, one can still speak of a spacetime structure at scales smaller than \eq{\ell}, it would be significantly different from that in the macroscopic domain --- ``mild'' deformations (such as, e.g., those accounting for spatial anisotropy~\cite{Sonego:2008iu}) are not enough, and one might well expect a radically new phenomenology~\cite{liberati-visser,hossenfelder-local,amelino-relloc,hossenfelder-soccer}.

The above argument can be rephrased in terms of symmetries of the physical laws. If (A)--(B), and (i)--(iv) all apply, we must expect strictly Poincar{\'e}-invariant laws, at any scale down to \eq{\ell_{\Rom{P}}}.  If, on the other hand, at least one of our postulates/assumptions breaks down around \eq{\ell}, the laws of physics will no longer be Poincar{\'e}-invariant at smaller scales.  If isotropy and the relativity principle are not affected, such symmetry breaking can only be severe --- not necessarily unviable, but potentially tightly constrained.

Remarkably, exact Poincar{\'e} invariance does not prevent \emph{a priori} the onset of new physics in the mesoscopic regime.  It is still possible that the very mechanisms leading to the sudden emergence of a classical spacetime end up introducing new Lorentz- and translation-invariant terms in the standard physical laws, suppressed by coefficients depending on powers of some \eq{\ell} --- hence, negligible at macroscopic scales (see e.g.,~\cite{Sorkin:2007qi}).

\bigskip

``Spacetime'' is a powerful word indeed. When seen as a ``grand stage'' for physical phenomena, we ought to either take it as is, or rebuild it from scratch, for it seems to offer no place for intermediate regimes. As a synonym for physical laws and their symmetries, however, it can provide room for a potentially rich, unforeseen phenomenology.

\bigskip\bigskip

\paragraph*{Acknowledgements} It is a pleasure to thank D. Benincasa and A. Belenchia for illuminating discussions.

%


\begin{thebibliography}{99}
%
{\small
%
\bibitem{grclassics}
S.~W.~Hawking and G.~F.~R.~Ellis, 
{\em The Large Scale Structure of Space-Time\/} 
(Cambridge University Press, Cambridge, 1973).
%
\bibitem{rovelli-loop}
C.~Rovelli,
  ``Loop quantum gravity: the first 25 years'', 
  \href{http://iopscience.iop.org/0264-9381/28/15/153002}{Class.~Quantum Grav.} {\bf 28} 153002 (2011),
  E-print \href{http://arxiv.org/abs/1012.4707}{arXiv:1012.4707 [gr-qc]}.
%
\bibitem{oriti-gft} 
D.~Oriti,
  ``The microscopic dynamics of quantum space as a group field theory'', 
  in J.~Murugan, A.~Weltman and G.~F.~R.~Ellis (eds.), {\em Foundations of Space and Time\/} (Cambridge University Press, Cambridge, 2012), pp.~257--320, 
  E-print \href{http://arxiv.org/abs/1110.5606}{arXiv:1110.5606 [hep-th]}. 
%
\bibitem{dowker-causets}
F.~Dowker,
  ``Introduction to causal sets and their phenomenology'',
  \href{http://link.springer.com/article/10.1007%2Fs10714-013-1569-y}{Gen.~Relativ.~Gravit.} {\bf 45}, 1651 (2013).
%
\bibitem{Sorkin:2007qi}
R.~D.~Sorkin,
  ``Does locality fail at intermediate length-scales?'',
  in D.~Oriti (ed.), {\em Approaches to Quantum Gravity\/} (Cambridge University Press, Cambridge, 2009), pp.~26--43,
  E-print \href{http://arxiv.org/abs/gr-qc/0703099}{gr-qc/0703099}.
%
\bibitem{Liberati:2013xla}
S.~Liberati,
  ``Tests of Lorentz invariance: a 2013 update'',
  \href{http://iopscience.iop.org/0264-9381/30/13/133001/article}{Class.~Quantum~Grav.} {\bf 30}, 133001 (2013),
  E-print \href{http://arxiv.org/abs/1304.5795}{arXiv:1304.5795 [gr-qc]}.
%
\bibitem{mtw}
C.~W.~Misner, K.~S.~Thorne and J.~A.~Wheeler, {\em Gravitation\/}  (Freeman, San Francisco, 1973).
%
\bibitem{brown2005}
H.~R.~Brown,
  {\em Physical Relativity\/} 
  (Oxford University Press, Oxford, 2005).
%
\bibitem{Sonego:2008iu}
S.~Sonego and M.~Pin,
  ``Foundations of anisotropic relativistic mechanics'',
  \href{http://scitation.aip.org/content/aip/journal/jmp/50/4/10.1063/1.3104065}{J.~Math.~Phys.} {\bf 50}, 042902 (2009),
  E-print \href{http://arxiv.org/abs/0812.1294}{arXiv:0812.1294 [gr-qc]}.
%
\bibitem{dray-signchange}
T.~Dray, C.~A.~Manogue, and R.~W.~Tucker, ``Particle production from signature change'', \href{http://link.springer.com/article/10.1007%2FBF00756915}
{Gen. Rel. Grav.} {\bf 23}, 967 (1991).
%
\bibitem{silke-signchange}
A.~White, S.~Weinfurtner, and M.~Visser, ``Signature change events: a challenge for quantum gravity?'',
  \href{http://iopscience.iop.org/0264-9381/27/4/045007/}{Class.\ Quant.\ Grav.\ } {\bf 27}, 045007 (2010), E-print \href{http://arxiv.org/abs/arXiv:0812.3744}{arXiv:0812.3744 [gr-qc]}.
%
\bibitem{Baccetti:2011aa}
V.~Baccetti, K.~Tate and M.~Visser,
  ``Inertial frames without the relativity principle'',
  \href{http://link.springer.com/article/10.1007%2FJHEP05(2012)119}{JHEP} {\bf 1205}, 119 (2012),
  E-print \href{http://arxiv.org/abs/1112.1466}{arXiv:1112.1466 [gr-qc]}.
%
\bibitem{Rovelli:2010ed}
C.~Rovelli and S.~Speziale,
  ``Lorentz covariance of loop quantum gravity'',
  \href{http://journals.aps.org/prd/abstract/10.1103/PhysRevD.83.104029}{Phys.~Rev.~D} {\bf 83}, 104029 (2011),
  E-print \href{http://arxiv.org/abs/1012.1739}{arXiv:1012.1739 [gr-qc]}.
%
\bibitem{Gibbons:2007iu} 
G.~W.~Gibbons, J.~Gomis and C.~N.~Pope,
  ``General very special relativity is Finsler geometry'', 
  \href{http://journals.aps.org/prd/abstract/10.1103/PhysRevD.76.081701}{Phys.~Rev.~D} {\bf 76}, 081701 (2007),
  E-print \href{http://arxiv.org/abs/0707.2174}{arXiv:0707.2174 [hep-th]}.
%
\bibitem{Hossdefects} 
S.~Hossenfelder,
  ``Phenomenology of space-time imperfection. I. Nonlocal defects,''
  \href{http://dx.doi.org/10.1103/PhysRevD.88.124030}{Phys.~Rev.~D} {\bf 88}, 124030 (2013), 
  E-print \href{http://arxiv.org/abs/1309.0311}{arXiv:1309.0311 [hep-ph]}.
%
\bibitem{liberati-visser}
S.~Liberati, S.~Sonego and M.~Visser,
  ``Interpreting doubly special relativity as a modified theory of measurement'',
  \href{http://journals.aps.org/prd/abstract/10.1103/PhysRevD.71.045001}{Phys.~Rev.~D} {\bf 71}, 045001 (2005), 
 E-print \href{http://arxiv.org/abs/gr-qc/0410113}{gr-qc/0410113}.
%
\bibitem{hossenfelder-local} 
S.~Hossenfelder,
  ``Bounds on an energy-dependent and observer-independent speed of light from violations of locality'', 
  \href{http://journals.aps.org/prl/abstract/10.1103/PhysRevLett.104.140402}{Phys.~Rev.~Lett.} {\bf 104}, 140402 (2010), 
  E-print \href{http://arxiv.org/abs/1004.0418}{arXiv:1004.0418 [hep-ph]}. 
%
\bibitem{amelino-relloc}
G.~Amelino-Camelia, L.~Freidel, J.~Kowalski-Glikman and L.~Smolin,
  ``The principle of relative locality'',
  \href{http://journals.aps.org/prd/abstract/10.1103/PhysRevD.84.084010}{Phys.~Rev.~D} {\bf 84}, 084010 (2011),
  E-print \href{http://arxiv.org/abs/1101.0931}{arXiv:1101.0931 [hep-th]}; 
%
\bibitem{hossenfelder-soccer}
S.~Hossenfelder,
  ``The soccer-ball problem'', 
  \href{http://mi.mathnet.ru/sigma939}{SIGMA} {\bf 10}, 074 (2014), 
  E-print \href{http://arxiv.org/abs/1403.2080}{arXiv:1403.2080 [gr-qc]}.
%
}
%
\end{thebibliography}
\end{document}